\documentclass[aps,prl,twocolumn,showpacs,groupaddress]{revtex4}
\usepackage{amsmath}
\usepackage{amsfonts}
\usepackage{amssymb}
\usepackage{epsfig}
\usepackage{graphicx}
\newcommand{\be}{\begin{equation}}
\newcommand{\ee}{\end{equation}}

\begin{document}

\title{Overlap fluctuations in glass-forming liquids} 

\author{Ludovic Berthier}
\affiliation{Laboratoire Charles Coulomb, UMR 5221, CNRS and Universit\'e
Montpellier 2, Montpellier, France}

\date{\today}

\begin{abstract}
We analyse numerically thermal fluctuations of the static overlap between 
equilibrium configurations in a glass-forming liquid 
approaching the glass transition. We find that the emergence of 
slow dynamics near the onset temperature 
correlates with the development of non-Gaussian 
probability distributions of overlap fluctuations, measured using 
both annealed and quenched definitions. Below a critical temperature, 
a thermodynamic field conjugate to the overlap induces a first-order 
phase transition, whose existence we numerically demonstrate in the annealed 
case. These results establish that the approach to the glass transition 
is accompanied by profound changes in the nature of thermodynamic fluctuations,
deconstructing the view that glassy dynamics occurs with little structural 
evolution.
\end{abstract}

\pacs{05.10.-a, 05.20.Jj, 64.70.Q-}




\maketitle

Theoretical approaches to the physics of glass-forming materials
are broadly organized in two categories~\cite{rmp}. A first class of theories 
concentrates on thermodynamic aspects, typically 
starting from a description of (assumed) relevant structural features of 
viscous liquids (configurational entropy, geometrical motifs, free volume), 
from which slow dynamics is predicted to emerge~\cite{walter}. 
A second class of models is based on the opposite view 
that the thermodynamics of viscous liquids is not evolving 
in any essential way, and focuses directly on relaxational aspects. 
This dynamical viewpoint is justified by the observation 
that the structure of viscous liquids does not seem to differ 
drastically from that of simple liquids, at least at the level of 
two-body static correlations. In recent years, this view gained 
support as it can directly be connected to detailed studies of dynamic 
heterogeneity in glassy materials~\cite{ediger}, 
which have unambiguously established that nontrivial spatio-temporal 
fluctuations accompany the glass transition~\cite{book}. 

In this work, we show that nontrivial, measurable 
thermodynamic fluctuations develop 
in supercooled liquids approaching the glass transition.
We characterize their nature and show that 
they are also intimately related to dynamics. 
To obtain these results, we analyze 
the thermal fluctuations of a thermodynamic quantity, the overlap 
between equilibrium configurations. The physical motivation
is that if the glass transition 
is controlled by a sharp decrease in the number of available 
metastable states~\cite{AG} possibly leading to the entropy crisis
first discussed by Kauzmann~\cite{kauzmann},
it should then become more likely for two independent 
equilibrium configurations to belong to the same state,   
and thus to have a large mutual overlap.  Therefore, 
thermal fluctuations of the overlap, just as the more 
technical construction of point-to-set correlations~\cite{BB},
should directly reveal and quantify the emergence of growing 
structural correlations in glass-forming liquids 
approaching the glass transition.

In the context of supercooled liquids, the fluctuations of the
overlap $Q$ between equilibrium configurations
have first been analyzed for spin glass models 
displaying a random first order transition (RFOT)~\cite{beyond,KT},
where the overlap distribution is needed to characterize the low-temperature 
phase. However, the overlap is also useful {\it above} the 
glass transition, as it allows the introduction 
of a Landau free energy $V(Q)$, also called 
`effective potential'~\cite{FP}. The potential 
was shown theoretically to 
capture the temperature evolution of RFOT free energy landscapes.
In the mean-field limit where these concepts are well-defined,  
$V(Q)$ loses convexity when metastable states first appear,
it then develops a local minimum at the mode-coupling singularity,
which becomes the global one at the `ideal' or Kauzmann
glass transition~\cite{FP}.
Direct measurements of $V(Q)$ in finite dimensions are scarce
and conflicting~\cite{coluzzi,mauro,parisi,giacomo}.
It was found to display none of the mean-field features 
in two lattice glass models~\cite{mauro,parisi}, 
while a recent investigation using soft spheres suggests 
a change in the convexity of 
$V(Q)$ near the mode-coupling temperature~\cite{giacomo}. 

The potential $V(Q)$ also serves as a starting point for field-theoretical 
calculations attempting to extend RFOT results 
to finite dimensions~\cite{moore,dzero,silvio2,gilles,silvio}. 
These calculations 
additionally suggest that the RFOT mean-field landscape is 
highly fragile with respect to finite dimensional 
fluctuations~\cite{fragile}, which could even affect 
the universality class to be considered~\cite{drossel}. 
Since these findings directly challenge the relevance of a
thermodynamic perspective to supercooled liquids, detailed studies 
of $V(Q)$ in finite dimensions are needed. 

A more direct interpretation of the effective potential is obtained 
from its definition as a `large deviation' function for the 
equilibrium fluctuations of the overlap, 
\be
P(Q) \sim \exp[ -\beta N V(Q) ],
\label{def}
\ee
where $P(Q)$ is the probability distribution of 
equilibrium overlap fluctuations in a system with $N$ particles
at temperature $T = \beta^{-1}$ (we set Boltzmann's constant to unity). 
Equation (\ref{def}) shows that the temperature evolution 
of $V(Q)$ directly affects the nature of 
thermal fluctuations of the overlap, and also suggests a 
conceptually simple way of measuring $V(Q)$. 
Interestingly, Eq.~(\ref{def}) provides a direct connection 
with dynamical views of glasses.
Recently, large deviations of 
{\it dynamical} observables have been analyzed~\cite{merolle}.
The emergence of spatially heterogeneous dynamics
was related to the appearance of non-Gaussian (nearly exponential) 
probability distributions of dynamic fluctuations.
Equivalently, these broad tails imply that a field 
conjugated to the dynamic activity should induce a nonequilibrium 
first-order phase transition between two phases with 
distinct dynamics, as observed numerically~\cite{rob_prl07,rob_science}.  
While the existence of long-lived metastable states 
(as in RFOT) is sufficient to explain these dynamic fluctuations and 
nonequilibrium transitions~\cite{rob_pspin}, alternative explanations with 
trivial thermodynamics also exist~\cite{rob_prl07}. 
Therefore, establishing the existence of thermodynamic observables 
obeying a phenomenology similar to dynamic ones 
will provide a concrete bridge between static and dynamic
viewpoints~\cite{paddy}.

We use computer simulations to analyze static 
fluctuations of the overlap in a simple numerical model of a 
glass-forming material.
We consider a 50:50 binary mixture of harmonic
spheres~\cite{berthier_09} of diameter ratio 1.4,
which we study using Monte Carlo dynamics.
The Hamiltonian reads $H( \{ {\bf r} \} )  = \sum_{j>i} 
v(\frac{|{\bf r}_{i} - {\bf r}_j|}{\sigma_{ij}})$, with the harmonic pair 
interaction $v (r \leq 1) = \frac{E}{2} (1-r)^2$, truncated for 
distances larger than the mean diameter $\sigma_{ij} = \frac{1}{2}
(\sigma_i + \sigma_j)$, and $\{ {\bf r} \} \equiv ({\bf r}_1, 
\cdots, {\bf r}_n)$.
For the density 
$\rho=0.675$ (using the small particle diameter as unit length),
this model behaves as a binary hard sphere mixture~\cite{berthier_09}, 
which is a well established model to analyze the glass transition. 
It is characterized by an 
onset temperature around $T_{\rm on} \approx 10$, and a mode-coupling
temperature $T_{\rm mct} \approx 5.2$, with 
temperatures expressed in units of $10^{-4} E$~\cite{kob_12}.
The overlap $Q_{12}$
between configurations $1$ and $2$ is defined as 
\be
Q_{12} = \frac{1}{N} \sum_{i,j=1}^N
\theta( a- | {\bf r}_{1,i} - {\bf r}_{2,j} | ),
\ee 
where $\theta(x)$ is the Heaviside function, 
${\bf r}_{1,i}$ denotes the position 
of particle $i$ within configuration 1, and we take $a=0.3$. 
By definition, $Q_{11} =1$, 
while $Q_{12}$ is small for uncorrelated configurations.  
(of order $\approx \frac{4}{3} \pi \rho a^3 
\equiv Q_{\rm rand} \ll 1$).  
Note that exchanging the positions 
of two particles does not decrease $Q_{12}$. Therefore, the overlap 
represents an `agnostic' measure of the degree of similarity between
two amorphous density fields, with no reference to a specific type
of structural order.

By definition, $V(Q)$ represents 
the free energy cost to maintain two thermalized copies of the 
liquid at a fixed value of their mutual overlap. 
Formally, this amounts to performing 
the following `{\it quenched}' calculation:
\be
V_{\rm q}(Q) = - \frac{T}{N} 
\int d{\bf r}_2 \frac{e^{-\beta H_2}}{Z_2} 
\log
\int d {\bf r}_1 {e^{-\beta H_1} } \delta (Q-Q_{12}),
\label{quenched}
\ee  
where $H_1 \equiv H (\{ {\bf r}_1 \} )$ and 
$H_2 \equiv H (\{ {\bf r}_2 \} )$, while $Z_2$ is 
the corresponding partition functions. In Eq.~(\ref{quenched}),
the thermal fluctuations of $Q_{12}$ 
are first probed for a fixed configuration 2 drawn from 
the equilibrium distribution, and then the logarithm of the 
probability distribution is averaged by sampling 
independent configurations. 

This procedure is numerically demanding as it requires 
two successive averages. A simpler, but approximate, 
procedure is to use an `{\it annealed}' definition:
\be
V_{\rm a}(Q) = - \frac{T}{N} \log \iint  d{\bf r}_2 
d {\bf r}_1  { e^{-\beta ( H_2 + H_1 )   } } 
\delta (Q-Q_{12}),
\label{annealed}
\ee  
where configurations 1 and 2 are fluctuating simultaneously,
and no disorder average is needed.
 
Direct measurements of $V(Q)$ are difficult because typical 
fluctuations of $Q$ are small compared to the average value 
$\approx Q_{\rm rand}$. To probe large deviations 
of the overlap, we use umbrella sampling techniques to measure the 
statistical weight of untypical values of the overlap. In practice, 
we use for each temperature $T$ a series of $n$ independent 
simulations, each simulation being biased by a 
Gaussian perturbation to the original Hamiltonian,  
$W_i(Q) = k_i (Q - Q_i)^2$, for $(i = 1, \cdots, n)$, 
which biases the overlap towards a desired value 
$Q_i \in [Q_{\rm rand}, 1]$. We make sure that each 
independent simulation first reaches the (biased) equilibrium,
and that simulations are long enough that they can 
properly sample equilibrium fluctuations in the 
biased phase space. Thus, each simulations returns the 
measurement of the (biased) probability distribution 
functions, $P_i(Q)$. We then use multi-histogram reweighting 
methods to reconstruct the unbiased probability $P(Q)$ 
from the $n$ independently measured $P_i(Q)$~\cite{frenkel}, 
\be
P(Q) = \frac{\sum_{i=1}^n P_i(Q) }{ \sum_{i=1}^n e^{-\beta W_i} / Z_i },
\label{pq}
\ee 
where the $Z_i$ are defined self-consistently as 
\be
Z_i = \int_0^1 dQ' \frac{ \sum_{j=1}^n P_j(Q') }{ \sum_{j=1}^n 
e^{\beta(W_i-W_j) }/Z_j }.
\ee 
We find that up to 16 independent simulations are needed
to accurately reconstruct $P(Q)$ over the entire 
relevant range, depending on the system size studied, 
$N=64$, $108$ and $256$, and on the temperature, $T \geq 7$.
We were not able to properly sample fluctuations for 
$T<7$ (and thus closer to $T_{\rm mct}$).
The more demanding simulations are for large $Q$, large $N$ 
and low $T$. Up to 40 independent samples were
used for the disorder average in Eq.~(\ref{quenched}). 
Finally, note that using biasing potentials $W_i(Q)$ 
efficiently solves the problem (first discussed in Ref.~\cite{coluzzi}) 
of translational and rotational invariances in Eqs.~(\ref{quenched}, 
\ref{annealed}).

\begin{figure}
\psfig{file=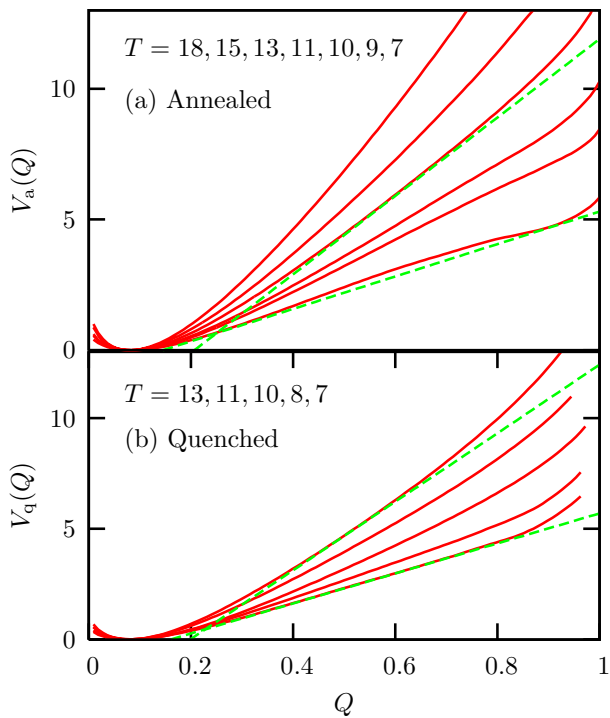,width=8.cm}
\caption{Temperature evolution of effective potential 
using (a) annealed and (b) quenched averages, for $N=108$ particles. 
Very similar behaviour is obtained for $N=64$ and 
$N=256$ (not shown). Data are vertically shifted data at different $T$
such that $V(Q)=0$ at the low-$Q$ mininum. Dashed lines 
represent straight lines, and temperature decreases from top to bottom.} 
\label{fig1}
\end{figure}

We present in Fig.~\ref{fig1} the numerical results obtained for both 
$V_{\rm q}(Q)$ and $V_{\rm a}(Q)$ 
in harmonic spheres for a range of temperatures, 
$T \geq 7$, which thus encompasses the onset of slow dynamics. 
These results indicate that thermal fluctuations of the overlap
become broader as temperature is lowered and deviate increasingly 
from a Gaussian behaviour, which would correspond, via Eq.~(\ref{def}),
to a parabolic $V(Q)$. As suggested by the dashed lines, 
the fluctuations are well described for temperatures $T \lesssim 10$ 
and for intermediate $Q$ by an exponential behaviour.
Note that for the annealed case at the lowest $T$
the potential is clearly not convex, at least for this 
moderate system size~\cite{maxwell}. 
Overall, this behaviour is in excellent agreement 
with results obtained within mean-field models displaying a RFOT, 
where the convexity of $V_{\rm q}(Q)$ is lost below $T_{\rm on}$. 
For finite dimensional systems, 
convexity should be restored through the emergence of 
interfaces and phase separation between high-$Q$ and low-$Q$ 
phases~\cite{giacomo}, therefore yielding exponential decay in $P(Q)$, 
and thus linear behaviour for $V(Q) \sim -T \log P(Q)$, as 
observed in Fig.~\ref{fig1}. We find quantitative,
rather than qualitative, differences between $V_{\rm a}$ and $V_{\rm q}$.
The main effect of the quenched disorder in these data
is to introduce an additional source of fluctuations 
which depresses slightly the emergence of 
exponential decay from $T \approx 10$ for the annealed case to 
$T \approx 8$ for the quenched case.
  
It is remarkable that $V(Q)$, which quantifies the 
thermal fluctuations of a {\it purely static} observable, 
loses convexity near (or slightly below) the onset 
temperature. Below $T_{\rm on}$, 
time correlation functions develop a two-step decay,
and dynamics become spatially  heterogeneous. Our results are thus 
qualitatively distinct from the emergence of non-Gaussian
fluctuations of dynamic observables~\cite{merolle}, and 
they demonstrate that thermodynamic fluctuations are (at least) as 
relevant as dynamic ones. 
The physical interpretation of the behaviour of $V(Q)$
offered by RFOT is that $T_{\rm on}$ marks the emergence 
of many metastable states, whose number decreases
as temperature is lowered further. This makes it more
and more likely for two configurations drawn at random to belong 
the same state and thus to have a large mutual overlap, 
as observed in Fig.~\ref{fig1}. This also 
suggests that the driving force for structural relaxation 
is reduced at low $T$, which is the RFOT theory explanation for the 
slowing down of the dynamics~\cite{rfot}. 

A direct, but spectacular, consequence 
of the loss of convexity of $V(Q)$ is that a field conjugated 
to the overlap should induce an {\it equilibrium} first-order 
phase transition~\cite{FP,jorge,marc,giorgio,cardenas}, 
because its main effect is to `tilt' 
the potential towards large $Q$ values. Physically, this amounts 
to studying the phase diagram of two coupled copies of the same system:
\be
H_{\rm tot} (\{ {\bf r}_1 \}, \{{\bf r}_2 \}) = 
H(\{ {\bf r}_1 \}) + H(\{ {\bf r}_2\}) - \epsilon Q_{12}.
\label{ham}
\ee
In the quenched version, copy 2 is drawn from the equilibrium distribution, 
the thermal properties of copy 1 are measured and then 
averaged over independent copies 2. In the annealed scheme, 
the copies evolve simultaneously under the influence of $H_{\rm tot}$
in Eq.~(\ref{ham}).
Generalizing Eq.~(\ref{pq}) to take into account the presence
of the thermodynamic field $\epsilon$, we directly estimate 
$P(Q,\epsilon)$ from the set of numerical 
simulations described above. We can then explore relevant features of 
the $(T, \epsilon)$ phase diagram.

\begin{figure}
\psfig{file=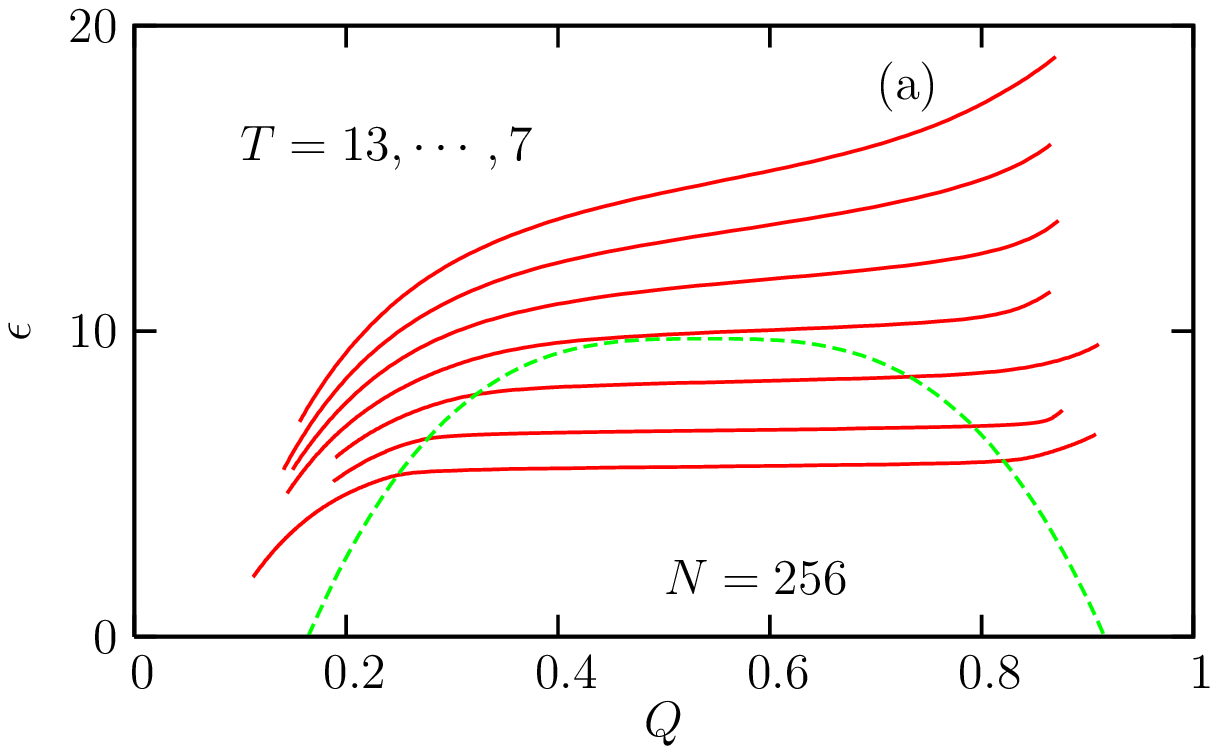,width=8.cm,clip}
\psfig{file=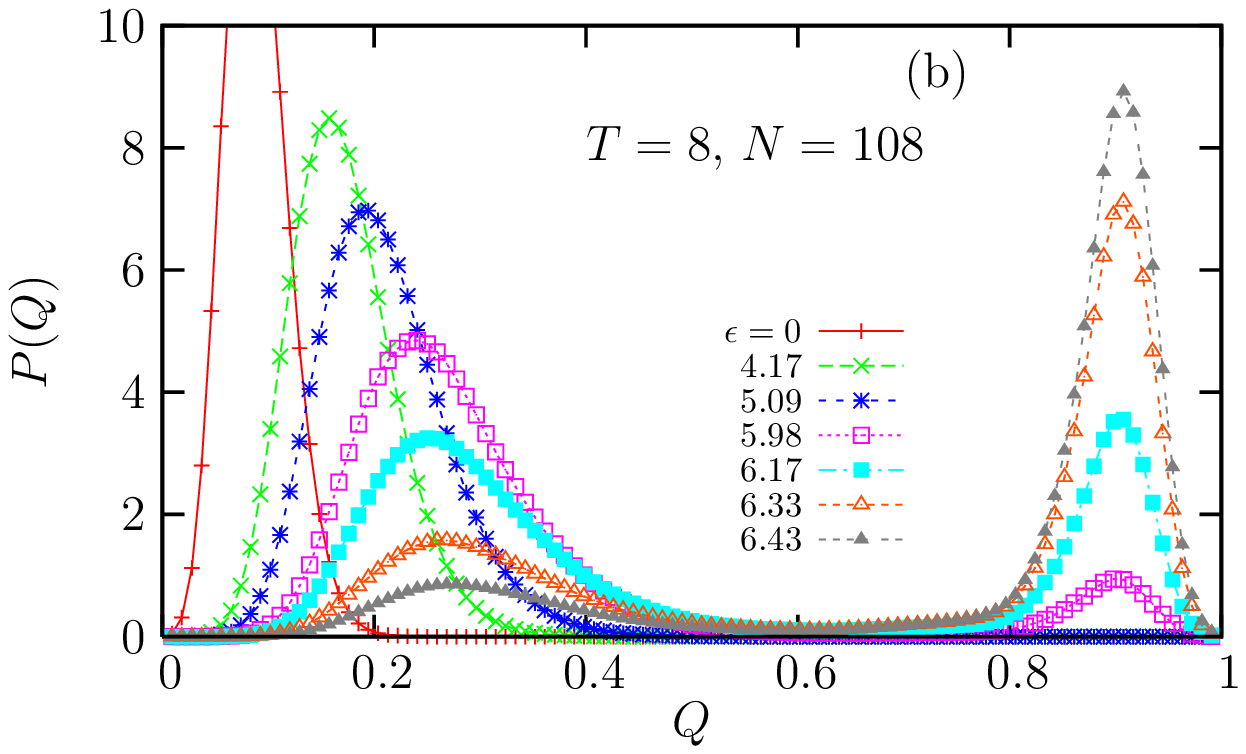,width=8.cm}
\psfig{file=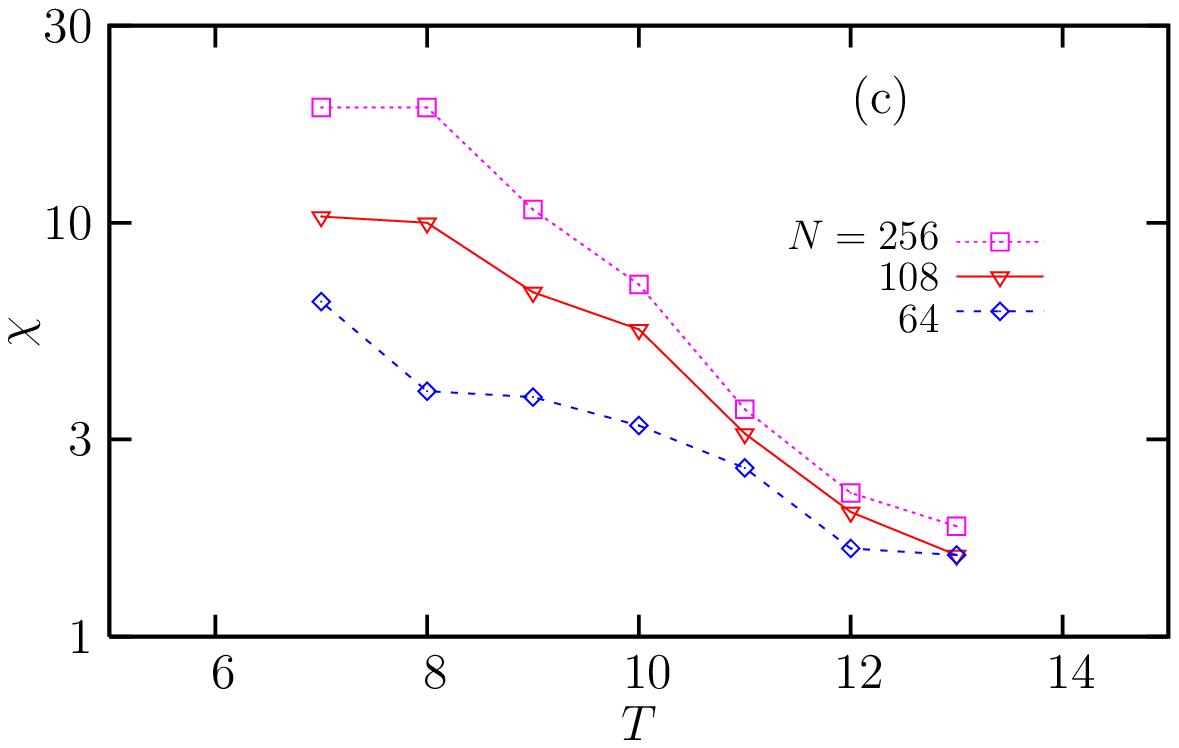,width=7.7cm}
\caption{Numerical indications of a 
thermodynamic first-order equilibrium phase transition
ending at a critical point near ($T_c \approx 9.8$, 
$\epsilon_c \approx 10$).
(a) Isotherms $\epsilon(Q)$ for $N=256$, the dashed coexistence line 
is drawn using the $d=3$ Ising model critical exponent. 
(b) Probability distribution of overlap fluctuations
across the first-order transition at $T=8$ and $N=108$, 
the coexistence occurring near $\epsilon \approx 6.17$. 
(c) Temperature evolution of the maximum of the static 
susceptibility for different system sizes.
} 
\label{fig2}
\end{figure}

We present in Fig.~\ref{fig2} our main findings for the annealed case, 
which establish the existence of first-order phase transition 
terminating at a second order critical point.  
Figure \ref{fig2}a shows the evolution of isotherms 
$\epsilon (Q)$, in a representation which underlies 
the analogy with the standard liquid-gas coexistence region. 
While $Q$ increases smoothly with 
$\epsilon$ at high temperature, it develops a sharp jump 
as temperature becomes lower than $T \approx 10$.  
By construction, this must correspond to the temperature where $V_{\rm a}(Q)$
loses convexity in Fig.~\ref{fig1}.
A stronger indication of the emergence of a first-order 
phase transition is obtained by measuring the fluctuations 
of the overlap at finite $\epsilon$, 
as shown in Fig.~\ref{fig2}b. While the
fluctuations are nearly Gaussian for small and large values of 
the coupling field, they are clearly {\it bimodal} at intermediate $\epsilon$, 
with peak positions revealing the values of the overlap in the 
coexisting two phases. Finally, Fig.~\ref{fig2}c 
presents data for the static susceptibility 
$\chi(Q,\epsilon) = N [ \langle Q^2 \rangle - \langle Q \rangle^2 ]$.
Increasing $\epsilon$ at constant $T$ we find that $\chi$ displays 
a maximum at a well-defined value of the field, 
which coincides with the value for which $P(Q)$ is bimodal. 
We report the temperature evolution of this maximum for various system sizes 
in Fig.~\ref{fig2}c. These data indicate that fluctuations are enhanced with 
increasing $N$ at low enough temperature, supporting the existence of 
a first-order phase transition in the thermodynamic limit 
below a critical temperature $T_c$, which is expected to be in the
same universality class as the $d=3$ Ising model~\cite{giulio2}. 
Indeed, our data are compatible with 
$\chi \sim L^d$ at low-$T$, while the data for $\chi / L^{\gamma/\nu}$
cross near $T_c \approx 9.8 \pm 1.$ when using the $3d$ Ising values 
of the critical exponents. We note that the isotherms 
in Fig.~\ref{fig2}a are well-described below $T_c$ by 
a jump in $Q$ increasing as $\Delta Q \sim 
(\epsilon_c- \epsilon)^\beta$ using again the Ising value for 
$\beta$ and $\epsilon_c \approx 10$. 
By contrast, the quenched coupling is believed 
to be in a different universality class, the one of the random 
field Ising model~\cite{giulio3}. 
We would need data at lower temperature 
to test this interesting prediction, a task we leave for future
work.  


The present results unambiguously demonstrate 
the emergence of strongly non-Gaussian thermodynamic fluctuations 
in a three-dimensional, 
bulk supercooled liquids approaching its glass transition. This is
also revealed by the existence, which we establish 
using finite size scaling analysis, of an equilibrium
first-order phase transition in the $(T, \epsilon)$. Such a 
phase transition was hinted in earlier numerical 
studies~\cite{giacomo,giorgio,cardenas}, but thermalization 
and sampling issues,
finite size effects, the location of the critical point and its 
connection with the onset of slow dynamics had not been discussed. 
  
This shows that the nature of $V(Q)$ in finite dimensional liquids
is compatible with the mean-field RFOT starting point used in 
field-theoretical calculations, and seems to contradict the claim 
that a different form of the potential should be 
used~\cite{drossel}. It also shows, somewhat surprisingly, 
that mean-field results are more robust for real liquids than 
for more abstract spin glass models~\cite{fragile}. 

Interestingly, the present first-order transition
is more easily studied numerically than the 
transition induced by a random pinning field 
recently analyzed for the same model~\cite{prl13}.
While both transitions result from the unique properties of 
RFOT free energy landscapes, only the latter corresponds to an ideal
glass transition line~\cite{giulio}, of the type 
possibly occurring in bulk liquids at low temperature.
It would be interesting to perform a finite size 
scaling analysis of the type presented here for 
the random pinning case as well. 

Although of purely thermodynamic origin, 
the present phase transition shares in fact many similarities 
with the nonequilibrium transition 
induced by a field conjugate to the 
dynamic activity~\cite{rob_prl07}. Both are first-order 
transitions induced by an external biasing field and   
differ qualitatively from the bulk glass transition.
Their qualitative similarity is further demonstrated
by the observation that the jump in the overlap 
reported in Fig.~\ref{fig2}a is accompanied by a sharp
change in the dynamics. We find for instance a decrease of 
3 decades of the self-diffusion constant for $T=9$ when 
the overlap jumps from $0.25$ to 0.7. This shows that 
a first-order change of the dynamic activity can in fact 
be easily triggered by a thermodynamic field in
fully equilibrium conditions. Combined to the results 
in Ref.~\cite{rob_pspin}, our work suggests that 
non-equilibrium first-order transitions in space-time 
are natural consequences of the emergence of a non-trivial 
effective potential $V(Q)$, which efficiently 
captures the complexity of the underlying
free energy landscape.

More generally, the parallel evolution of 
static and dynamic fluctuations unveiled here 
suggests that the temperature evolution of thermodynamic fluctuations 
drives the slow dynamics in glass-forming liquids,
deconstructing the familiar view that glassy dynamics occurs with 
little structural evolution.

\acknowledgments 
I thank D. Coslovich for helping me set up the numerical  
simulations and several useful discussions, 
and G. Parisi for pointing out an inconsistency in 
temperature units in an earlier version 
of the manuscript.
I also thank  G. Biroli, S. Franz, 
R. Jack, J. Kurchan, G. Szamel, G. Tarjus, and F. Zamponi for
additional exchanges.
The research leading to these results has received funding
from the European Research Council under the European Union's Seventh
Framework Programme (FP7/2007-2013) / ERC Grant agreement No 306845.


\begin{thebibliography}{99}
\bibitem{maxwell}
As for ordinary first-order phase transitions,
a non-convex potential is obtained in the mean-field limit 
which only considers homogeneous systems. In finite dimensions, 
the potential becomes convex because the system creates 
an interface between the two phases, which restores
convexity in the thermodynamic limit. 
The non-convexity found numerically here is therefore 
a finite-size effect, not a direct signature of the 
mean-field result.
\end{thebibliography}
\end{document}